\documentstyle[11pt,aaspp4,epsfig]{article}
\received{}
\accepted{November 19, 1999}
\slugcomment{To appear in ApJ.L}
\def\Epk{E_{\rm pk}}
\def\Epkmax{E_{\rm pk,0}}

\def\NEt{N_{\rm E}(E,t)}
\def\E0{E_{\rm 0}}
\def\P0{\Phi_{\rm 0}}
\def\Ep0{E_{\rm p0}}

\def\N0{N_{\rm 0}}
\def\t0{\t_{\rm 0}}

\def\Gt{{\bf G}(t)}
\def\Gtt{{\bf G}(t;\tau,\delta)}
\newcommand{\ltsima} {$\; \buildrel < \over \sim \;$}
\newcommand{\gtsima} {$\; \buildrel > \over \sim \;$}
\newcommand{\lta} {\lower.5ex\hbox{\ltsima}}
\newcommand{\gta} {\lower.5ex\hbox{\gtsima}}

\lefthead{Ryde \& Svensson}
\righthead{Time Evolution of GRB pulses}

\begin{document}

\title{On the Time Evolution of Gamma-Ray Burst
Pulses: A Self-Consistent Description}
\author{Felix Ryde and Roland Svensson\altaffilmark{1}}
\affil{Stockholm Observatory, SE-133 36 Saltsj\"obaden, Sweden}

\altaffiltext{1}{Institute of Theoretical Physics, UCSB, Santa Barbara
CA 93106, USA} 

\begin{abstract}
For the first time, the consequences of {\it combining} two well-established
empirical relations, describing different aspects of the spectral
evolution of observed
gamma-ray burst (GRB) pulses, are explored. These
empirical relations are: i) the hardness-intensity correlation,  and
ii) the hardness-photon fluence correlation. From these we find a 
self-consistent, quantitative, and compact description for the temporal
evolution of pulse {\it decay} phases within a GRB light curve.
In particular, we show that in the case of the two empirical
relations both being valid, the instantaneous photon flux
(intensity) must behave as $1/(1+ t/\tau)$ where $\tau$ is a time
constant that can be expressed in terms of the parameters of the two
empirical relations. The time evolution is fully
defined by two initial constants, and two parameters. 
We study a complete sample of 83 bright GRB pulses 
observed by the {\it Compton Gamma-Ray Observatory} and 
identify a major subgroup of GRB pulses ($\sim 
45 \%$), which satisfy the spectral-temporal behavior described above. 
In particular, the decay phase follows a reciprocal law in time.
It is unclear what physics causes such a decay phase.
\end{abstract}

\keywords{gamma rays: bursts}

\section{Introduction}

The mechanisms giving rise to the observed gamma-ray burst (GRB) emission
may reveal themselves in correlations describing the {\it continuum spectral
evolution}.
A few such correlations between observable quantities have been found.
One of these is that between 
the instantaneous hardness of the spectrum and the instantaneous
flux within
individual pulses, especially during their decay phases.
Somewhat more than every second pulse decay exhibits such a 
correlation (e.g., Kargatis et~al. 1995).
Furthermore, a correlation between the instantaneous 
hardness of the spectrum and
the  time-integrated flux, the fluence, has been established
in a majority of the pulses where it has been searched for (e.g., Crider
et~al. 1999). 

In this work, we demonstrate that for the decay phases of the light curve, 
for which both these
correlations are valid, the light curve will follow a specific decay law.
This is shown analytically in \S  2 by combining the two,
previously well-studied, empirical relations 
into a new compact description of the temporal
behavior of the  decay phase.
In \S  3, we study a sample of pulse decays 
observed by  the Burst and Transient Source 
Experiment (BATSE) on the {\it Compton Gamma-Ray Observatory (CGRO)}
and give a few illustrative examples. A discussion is given in \S 4.

\section{Descriptions of the Time Evolution}

The instantaneous photon spectrum, $\NEt$
[photons cm$^{-2}$ s$^{-1}$ keV$^{-1}$], having approximately the shape of a
broken power-law, is  characterized mainly by two entities, the total
instantaneous photon flux, $N(t)$ [photons cm$^{-2}$ s$^{-1}$],
and a measure of the ``hardness'', e.g., the instantaneous
peak energy, $\Epk (t)$, of the $E^2\NEt$ spectrum. The time evolution
of, for instance, a pulse decay phase in a GRB light curve
can then be described by a vector function 
${\bf G}(t'-t_{\rm 0};{\rm parameters}) = \left(N(t'-t_{\rm 0}),
\Epk(t'-t_{\rm 0})\right)$,
where $t'$ is the time parameter and $t_{\rm 0}$ is the starting time of
the pulse decay. Apart from the running time parameter there are a
number
of parameters specific to the pulse decay. The initial value
at $t \equiv t'-t_{\rm 0}= 0$ is ${\bf G}(0) \equiv (\N0,\Epkmax)$.
The relation between the evolution of the instantaneous spectral
characteristics, e.g., $\Epk(t)$, and the corresponding intensity 
of the light curve, e.g., $N(t)$,
has been widely studied, leading to empirical relations 
describing the observed behavior within a GRB and even within 
individual pulses. The most common trend is
the hard-to-soft evolution, in which the peak energy of the spectrum
decreases monotonically over the entire pulse (Norris et~al. 1986).
A less common trend, where the hardness and the intensity
track each other, was found by 
Golenetskii et~al. (1983). Similar results were also reported
by Kargatis et~al. (1994) and Bhat et~al. (1994).
Especially for the decay phase of pulses, such a hardness-intensity
correlation (HIC) is common. Kargatis et al. (1995) found a strong
correlation for 28 pulse decay phases in a sample of
26 GRBs with pulse pairs. For the decay phase, the HIC can be expressed as
\begin{equation}
\Epk(t) = \Epkmax\left[\frac{N(t)}{\N0}\right]^{\delta},
\end{equation}
\noindent
where $\delta$ is the correlation index. 

A second empirical relation was found by Liang \& Kargatis
(1996), who pointed out a correlation between the peak energy
and the time-integrated photon flux,
\begin{equation}
E_{\rm pk} (t) = \Epkmax e ^{-\Phi (t) / \Phi_{\rm 0}},
\end{equation}
\noindent
where $\Phi (t)$ is the
photon fluence $\equiv \int _{0} ^{t} N(t'') \,dt'' $ [cm$^{-2}$] 
integrated from the time of $\Epkmax$
and $\Phi _{\rm 0}$ is the exponential decay
constant.  
In their discovery paper, Liang \& Kargatis (1996)
showed the correlation to be valid for several long, smooth, single
pulses and especially for their decays. 
They found that 35 out of 37 pulses were consistent or
marginally consistent with the correlation. This
work was followed by a series of papers by Crider and coworkers
(1997, 1998a, b, 1999). In Crider et~al. (1999), a sample of 41 pulses
in 26 GRBs were studied in greater detail confirming the
original discovery.
They, however, preferred a slight modification for the
description of the decay and studied the peak energy 
versus the energy fluence, instead. 
The two approaches are very similar and
do not describe fundamentally different trends of the decay.
This empirical relation was also studied
in Ryde \& Svensson (1999), who showed that the relation
can be used to  
derive the shape of the time-integrated spectrum, as this is the
result of integrating the evolving time-resolved spectra. 

The two relations given by equations (1) and (2) fully describe the
evolution. If these two relations are fulfilled,
one can show (e.g., in \S 4 below) that the 
function $\Gt = (N(t),\Epk (t))$ is given by
\begin{eqnarray}
N(t) & = & \frac{\N0}{(1 + t/\tau)},\\
\Epk(t) & = & \frac{\Epkmax}{(1 + t/\tau)^\delta},
\end{eqnarray}
\noindent
i.e., the instantaneous photon flux is a reciprocal function in time
with the time constant $\tau$.
The peak energy has a similar dependence, differing by the HIC
index $\delta$. The number of parameters in the present formulation is
limited to two, $\tau$ and $\delta$.

{}From equations (3) and (4), all the empirical results discussed above
follow. Eliminating the
explicit time dependence by combining the two
equations, the hardness-intensity correlation described by equation (1)
is found: $\Epk=\Epkmax(N/\N0)^\delta$. 
The function $\Gtt$ describes
this relation as 
a path in the $N-E_{\rm pk}$ plane, and as $N(t)$ and $\Epk(t)$ 
evolve in the same manner, except for the
exponent $\delta$, the relations (3) and (4) give rise to the
Golenetskii power-law (Eq. 1).
Furthermore, the photon fluence is found by integrating 
equation (3):
\begin{equation}
\Phi(t) = \N0 \tau \ln(1 + t / \tau),
\end{equation}
\noindent
which, when used to eliminate the $(1 + t / \tau)$-dependence in equation
(4), gives the hardness-fluence relation:
\begin{equation}
\Epk(t) = \Epkmax e^{- \delta \Phi(t) /\N0 \tau}.
\end{equation}
\noindent
Identifying this equation with equation (2), one finds
the exponential decay constant to be given by $\P0 \equiv \N0 \tau /
\delta $,
and thus that the time constant, 
\begin{equation}
\tau = \delta \P0/\N0.
\end{equation}
This is the crucial relation that connects the pulse timescale with
the properties of relations (1) and (2).

\section{BATSE Observations}

To verify and illustrate the results above, we searched for this 
specific spectral-temporal behavior in GRBs observed by BATSE.
We studied the high energy resolution Large Area Detector (LAD)
observations of GRBs. We selected bursts from
the BATSE catalog\footnote{
The BATSE GRB catalog is available online at:
http://gammaray.msfc.nasa.gov/batse/grb/data/catalog/} up to GRB 990126
(with trigger number 7353), with a peak flux  (50-300 keV on a 256 ms 
timescale)
greater than $5$ photons s$^{-1}$ cm$^{-2}$, giving totally 155 bursts having
useful LAD high energy resolution burst (HERB) data.
Out of these, we found
a sample of 59 GRBs with a total of 83 pulse decays, strong
enough to allow spectral fitting to be done of the time-resolved
spectra in at least four time bins with a signal-to-noise (S/N) ratio of 
$\gta 30$ in the $25-1900$ keV band. The purpose of the fitting is to
determine the hardness parameter, $E_{\rm pk}(t)$, and to allow the
deconvolution of the count spectrum in order to obtain $N(t)$. For
this, a larger S/N ratio is not motivated.

Commonly, the Large Area Detector (LAD) and its
4 energy-channel DISCSC data are used, and
light curves are studied in narrow spectral ranges,
typically $50-300$ keV, in units of  {\it count rates}.
For our analysis we use the LAD HERB data, since they have the
necessary higher energy resolution,
with 128 energy channels, and cover the maximal possible energy range, 
$25$ keV $- 1.9$ MeV. Furthermore, we use light curves in terms of {\it
photon flux} rather than {\it count rates}. 
The spectral fitting was done 
using the Band et~al. (1993) function, with all its parameters free.

Out of the 83 pulse decays in our sample, we found 38 ($\sim 45 \% $) to be
consistent with the reciprocal decay law in Equation (3). 
We start by presenting detailed results of the analysis of two of these
pulses, namely GRB 921207 (\# 2083)  and
GRB 950624, (\# 3648), see Table 1. A fit of the function
$N = N(t)$ (Eq. 3) to the data, gives the photon flux at $t=0$,
$\N0$, and  the time constant, $\tau$. Freezing $\tau$ to 
this value, the fit of $ \Epk(t)$ (Eq. 4) gives  the initial value of the 
peak energy, $\Epkmax$ 
and the HIC index $\delta$. 
The results of these fittings are given in the upper half of Table 1. 
The same information can
be found from fits of the empirical relations described by equations (1)
and (2). A fit of the hardness-fluence correlation (2) allows the 
determination of
$\Epkmax$ and the exponential decay constant, $\P0$. Freezing $\Epkmax$
to the value obtained, the fit of the hardness-intensity correlation (Eq. 1)
gives values for $N_{\rm 0}$ and $\delta$. These fitted values, 
as well as the computed values for $\tau = \delta \P0/\N0$,
are given in the lower half of Table 1. Note the excellent consistency 
between the two sets of fitted
parameters. In the data for the first pulse of GRB 921207, especially
regarding the dependence of $N(t)$ on $t$ and on $\Epk$, 
we do not find any strong indication for it being a double pulse,
as suggested by Crider et~al. (1998a) (See Fig. 1).

The results of the analysis of a few further
examples are given in Table 2. The fitted decay
phases of these are presented in Figure 1, where the 
linear $1/N(t)$ functions are displayed.  
The four parameters
describing the decay phases, $\N0$, $\Epkmax$, $\tau$, $\delta$, are
found from the fits of $\Gtt$ given by equations (3) and (4), 
while $\Phi_{\rm 0}$ is found from fits of the empirical relation (2). 
Equation (7) gives consistent $\tau$-values within the errors.
For our purposes, we need as high time-resolution as possible, but
still permitting proper spectral fitting. We therefore chose a
S/N$=30$. When possible, we redid the analysis with 
S/N$=45$, checking that we arrived at the same results, 
now, however, with lower temporal resolution.

The panel furthest to the right
in the lower row of Figure 1 illustrates a case,
GRB 960807 (\# 5567), for which the reciprocal law is not valid. In
this specific case, the hardness-fluence correlation (Eq.2)
is valid while the HIC is not a power
law. A detailed discussion of such cases is given in Ryde et~al. (1999).

\section{Discussion and Conclusions}

Equation (3) is an important result, as it
describes  how the intensity declines with time in the decay phase of a
GRB pulse, belonging to the subgroup of pulses for which both the two  
empirical relations, equations (1) 
and (2), are valid. 
The reciprocal, instantaneous intensity is a linear function in time,
$1/N(t) = \left(1 + t / \tau \right)/\N0$. 
This should be compared to the generally discussed
exponential decay, e.g., in the terms of a FRED (fast rise, {\it
exponential} or {\it stretched exponential} decay) often used to characterize
single pulses within GRBs (e.g. Norris et al., 1996).
Our result is an analytical result following from the two
empirical relations.
We note that $N(t)$ approaches $(1-t/ \tau) \N0$ for $t \ll \tau$, which
is the same time behavior as that of the stretched exponential
$N(t)= \N0{\rm{exp}}\left[-(t/\tau_*)^{\nu}\right]$ in the same limit. 
When $t \lta \nu \tau_*$, 
it is difficult to distinguish between the two  behaviors.
The hardness, represented by the peak energy, $E_{\rm pk}(t)$,
also declines reciprocally with time, but is stretched by the
$\delta$-power (see Eq. 4). A comparison between the intensity
decline and
the $\Epk$-decline is shown, e.g., in Ford et~al. (1995).

As shown by equation (5), the fluence increases logarithmically with
time. This divergent behavior must eventually
change, when the emission of radiation changes behavior, terminates,
or shifts out of the observed spectral range.

The differential equations governing the time evolution of $\Gtt$
are readily found.
Differentiating equation (2) gives
\begin{equation}
-\frac{d\Epk(t)}{dt}=\frac{\delta}{\tau \N0} N(t)  \Epk (t),
\end{equation}
\noindent
which, combined with equation (1), gives the equation for $\Epk(t)$ as
\begin{equation}
-\frac{d\Epk(t)}{dt}=\frac{\delta}{\tau (\Epkmax)^{1/\delta}}\Epk^{1+1/\delta}(t).
\end{equation}
\noindent
Furthermore, combining equations (1) and (2) to
\begin{equation}
N(t)=\N0 e^{-\Phi (t)/ \N0 \tau},
\end{equation}
\noindent
gives, after differentiation, the equation for $N(t)$ as
\begin{equation}
-\frac{dN(t)}{dt}=\frac{1}{\N0 \tau} N^2(t).
\end{equation}
\noindent
Integrating equations (9) and (11) then gives equations (4) and (3).

The description is complete. A different $N(t)$-shape of the decay phase will
of necessity lead to either that the hardness-intensity correlation or
the hardness-fluence correlation or both must have a different shape
from the well-observed empirical ones (Eqs. 1 and 2). 
Details are discussed in Ryde et~al. (1999).           

The description does, unfortunately, not uniquely point to a
specific radiation process. For instance, equation (8) could be due
to any thermal process. Consequently, it is consistent with saturated 
Comptonization, without extra heating, which though, in general, is
difficult to achieve. Such a scenario has,
however, been discussed by Liang \& Kargatis (1996) and Liang (1997).
In their theoretical, saturated Compton cooling model, Liang et~al.
(1997) arrive at the equation 
$N(t) = d\Phi/dt = k t ^s {\rm exp}(-n\Phi/\P0) (1+ \Epk(t))^n/b(t)$
(obtained by rewriting their Eq. 5), where $k$ is a constant and 
$b(t)$ =  (the soft photon injection rate/BATSE photon flux).
In order to simplify this equation, they set, 
 in a rather ad hoc manner (guided by
Monte Carlo simulations) 
$b(t) = (1+ \Epk(t))^n$,  resulting in the last two factors cancelling.
The resulting equation for $\Phi$ then becomes identical to
our equation (10) for the case $s=0$ (i.e., no Thomson thinning) and taking
$n= 1/\delta$. The resulting solutions are, of course, identical. We strongly
emphasize the difference. We obtain our results based on two empirical
relations, while  Liang et~al. obtain their results by
employing a rather ad hoc simplification within a theoretical model.

In conclusion, we find a subgroup ($\sim 45 \%$) of GRB 
pulse decays which behave in a similar
way, with the decay phases being reciprocal 
functions, Eqs. (3) and (4).
This should thus represent a signature of the underlying physical processes
giving rise to these pulses.
We have illustrated the results both analytically and by studying 
BATSE pulses.
The time evolution is fully defined by the two initial conditions
at the start of the decay, $\N0$ and $\Epkmax$, 
and by two parameters specific to the GRB pulse decay, $\tau$
and $\delta$, or, equivalently, $\P0$ and $\delta$.

\acknowledgments

This research made use of data obtained through
 the HEASARC Online Service provided by NASA/GSFC.
We are also grateful to the GROSSC for support.
We thank S. Larsson, L. Borgonovo, R. Preece, A. Beloborodov, and J. Poutanen
for useful discussions. 
We acknowledge support from the Swedish Natural Science Research
Council (NFR), the Anna-Greta and Holger Crafoord Fund, a NORDITA
grant, and NSF Grant No. PHY94-07194.

\newpage 
\begin{figure}
\centerline{\psfig{file=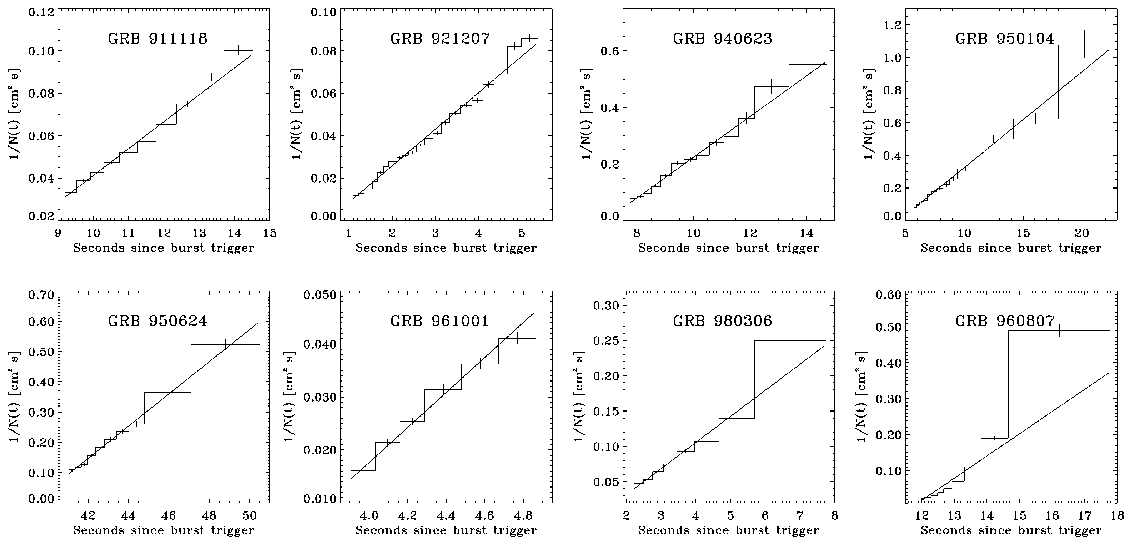, width=20cm}}
\caption{Observed, background-subtracted, photon flux
light curves of the decay phases for a few of the pulses in 
the sample studied. 
They are presented as $1/N(t)$ to clearly emphasize the
reciprocally linear behavior arrived at in equation (3).
The LAD (HERB) data are used and are rebinned to a S/N=$30$.
Fittings to the data give the parameters $\N0$ and
$\tau$. Fittings to the time evolution of $E_{\rm pk}(t)$ give,
correspondingly, 
$\Epkmax$ and $\delta$ by freezing $\tau$. Over a time $\tau$, the 
light curve decreases by a half.
GRB 960807 is shown as an example of a case which does not follow the 
spectral-temporal behavior discussed in the text. 
Note that the time is counted from the trigger time of the burst.
The time variable,
$t$, in the text has its zero-point at $t_{\rm 0}$, where
${\bf G}(0)= (N_{\rm 0} , \Epkmax)$.
}
\end{figure}

\newpage 

\begin{center}
\begin{deluxetable}{lll}
\footnotesize
\tablecaption{Fitted parameters for two pulses. \label{tbl-1}}
\tablewidth{0pt}
\tablehead{
\colhead{Parameters} & \colhead{\# 2083}   & \colhead{\# 3648}   
}
 
\startdata
$N_{\rm 0}$ [cm$^{-2}$ s$^{-1}$]\tablenotemark{a} & $99.5 \pm 1.3 $ &
$10.5\pm0.2$\nl
$\tau$ [s]\tablenotemark{a} & $0.58 \pm 0.01$ & $1.80\pm 0.08 $ \nl
$\Epkmax$ [keV]\tablenotemark{b}& $474\pm 9$ & $ 300 \pm 15$ \nl
$\delta$\tablenotemark{b} & $0.90\pm 0.02$& $1.07\pm0.05 $ \nl
\hline

$N_{\rm 0}$ [cm$^{-2}$ s$^{-1}$]\tablenotemark{d} & $100.4\pm1.2$ &
$10.6\pm0.3$ \nl
$\P0$ [cm$^{-2}$]\tablenotemark{c}  & $64.1\pm 1.5 $ & $16.9\pm0.9 $ \nl
$\Epkmax$ [keV]\tablenotemark{c}& $478\pm13$ & $310\pm18$ \nl
$\delta$\tablenotemark{d} & $0.89\pm 0.02 $ & $1.1\pm0.1 $ \nl
$\tau = \delta \P0/N_{\rm 0}$ [s] & $0.57\pm0.02$  & $1.75\pm0.19$ \nl
\enddata

\tablecomments{
{}$^{a}$Using Eq. (3).
{}$^{b}$Using Eq. (4) freezing $\tau$.
{}$^{c}$Using Eq. (2).
{}$^{d}$Using Eq. (1) freezing $\Epkmax$.
}

\end{deluxetable}
\end{center}

\begin{deluxetable}{lllllll}
\footnotesize
\tablecaption{The studied decay phases of  a few bright BATSE
GRB pulses \label{tbl-2}}
\tablewidth{0pt}
\tablehead{ 
\colhead{Burst Name} & \colhead{$N_{\rm 0}$}   & \colhead{$E_{\rm pk,0}$} & \colhead{$\tau$ }
 & \colhead{$\delta$}&  \colhead{$\Phi_{\rm0}$\tablenotemark{a}} &
\colhead{Time interval\tablenotemark{b}} \nl
 &\colhead{[cm$^{-2}$ s$^{-1}$]} & \colhead{[keV]} & \colhead{[s]}&  &
\colhead{[cm$^{-2}$]} & \colhead{[s]}  
}

\startdata
GRB 911118 (\#1085) & $32.1  \pm 0.4$ & $123.2\pm 2.4 $&$2.46 \pm 0.07$
 & $0.63 \pm 0.03 $ & $ 128\pm7 $ & $9.22-14.53$  \nl
GRB 921207 (\#2083) & $99.5 \pm 1.3$ & $474\pm 9 $&$ 0.58 \pm 0.01$
 & $0.90 \pm 0.02$ & $ 64.1\pm1.5 $ & $1.09-5.38$ \nl
GRB 940623 (\#3042) & $ 16.3 \pm 0.4$ & $523\pm 68 $ &$ 0.85 \pm 0.04$
 & $0.78 \pm 0.09$ & $ 18.4 \pm 2.6 $ & $7.74-14.72$  \nl
GRB 950104 (\#3345) & $14.0 \pm 0.4$ & $ 140\pm 18 $ &$ 1.22 \pm 0.07$
 & $0.41 \pm 0.10$ & $ 36 \pm 9 $ & $5.70-22.46 $  \nl
GRB 950624 (\#3648) & $10.5 \pm 0.2$ & $300\pm 15 $ &$ 1.80 \pm 0.08$
 & $1.07 \pm 0.05$ & $ 16.9 \pm 0.9 $ & $ 41.02-50.50   $  \nl
GRB 961001 (\#5621) & $ 69.1 \pm 1.8$ & $440\pm 60 $&$0.44 \pm 0.03$
 & $ 1.1\pm 0.2$ & $ 26 \pm 5  $ & $3.90-4.86$  \nl
GRB 980306 (\#6630) & $24.7 \pm 0.6$ & $253\pm 8 $ &$1.10 \pm 0.04$
 & $1.01 \pm 0.04$ & $ 27.9 \pm 1.2$ & $2.24-7.74 $  \nl
\enddata

\tablecomments{
{}$^a$As fitted from Eq. (2). 
{}$^b$Time interval during which the decay was studied.
}
\end{deluxetable}

\end{document}